\documentclass[prx, letterpaper, superscriptaddress, amsmath, amssymb, amssymb, reprint, floatfix]{revtex4-2}

\usepackage{graphicx}
\usepackage{dcolumn}
\usepackage{bm}
\usepackage{xcolor}
\usepackage{multirow}
\usepackage{upgreek}
\newcommand\pdfmath[1]{\texorpdfstring{$#1$}{#1}}

\usepackage{textcomp}
\usepackage{xcite}
\usepackage{verbatim}
\usepackage[colorinlistoftodos]{todonotes}
\usepackage{microtype}

\usepackage{xr-hyper}
\usepackage{hyperref}
\makeatletter
\newcommand*{\addFileDependency}[1]{
  \typeout{(#1)}
  \@addtofilelist{#1}
  \IfFileExists{#1}{}{\typeout{No file #1.}}
}
\makeatother

\newcommand{\appropto}{\mathrel{\vcenter{
  \offinterlineskip\halign{\hfil$##$\cr
    \propto\cr\noalign{\kern2pt}\sim\cr\noalign{\kern-2pt}}}}}

\begin{document}

\preprint{APS/123-QED}

\title{A quantum dot-based frequency multiplier}
\author{G. A. Oakes}
\email{These authors contributed equally to this work}
\affiliation{Cavendish Laboratory, University of Cambridge, J.J. Thomson Avenue, Cambridge CB3 0HE, United Kingdom}
\affiliation{Quantum Motion, 9 Sterling Way, London N7 9HJ, United Kingdom}
\author{L. Peri}
\email{These authors contributed equally to this work}
\affiliation{Cavendish Laboratory, University of Cambridge, J.J. Thomson Avenue, Cambridge CB3 0HE, United Kingdom}
\affiliation{Quantum Motion, 9 Sterling Way, London N7 9HJ, United Kingdom}
\author{L. Cochrane}
\affiliation{Cavendish Laboratory, University of Cambridge, J.J. Thomson Avenue, Cambridge CB3 0HE, United Kingdom}
\affiliation{Quantum Motion, 9 Sterling Way, London N7 9HJ, United Kingdom}
 \author{F.~Martins}%
 \affiliation{Hitachi Cambridge Laboratory, J.J. Thomson Avenue, Cambridge CB3 0HE, United Kingdom}
\author{L. Hutin}%
 \affiliation{CEA, LETI, Minatec Campus, F-38054 Grenoble, France}
\author{B. Bertrand}%
 \affiliation{CEA, LETI, Minatec Campus, F-38054 Grenoble, France}
\author{M. Vinet}%
 \affiliation{CEA, LETI, Minatec Campus, F-38054 Grenoble, France}
\author{A. Gomez Saiz}
 \affiliation{Quantum Motion, 9 Sterling Way, London N7 9HJ, United Kingdom} 
\author{C. J. B. Ford}
 \affiliation{Cavendish Laboratory, University of Cambridge, J.J. Thomson Avenue, Cambridge CB3 0HE, United Kingdom}
\author{C. G. Smith}
 \affiliation{Cavendish Laboratory, University of Cambridge, J.J. Thomson Avenue, Cambridge CB3 0HE, United Kingdom}
 \affiliation{Hitachi Cambridge Laboratory, J.J. Thomson Avenue, Cambridge CB3 0HE, United Kingdom}
\author{M. F. Gonzalez-Zalba}
 \email{fernando@quantummotion.tech}
 \affiliation{Quantum Motion, 9 Sterling Way, London N7 9HJ, United Kingdom}

\date{\today}

\begin{abstract}
 Silicon offers the enticing opportunity to integrate hybrid quantum-classical computing systems on a single platform. For qubit control and readout, high-frequency signals are required. Therefore, devices that can facilitate its generation are needed. Here, we present a quantum dot-based radiofrequency multiplier operated at cryogenic temperatures. The device is based on the non-linear capacitance-voltage characteristics of quantum dot systems arising from their low-dimensional density of states. We implement the multiplier in a multi-gate silicon nanowire transistor using two complementary device configurations: a single quantum dot coupled to a charge reservoir and a coupled double quantum dot. We study the harmonic voltage conversion as a function of energy detuning, multiplication factor and harmonic phase noise and find near ideal performance up to a multiplication factor of 10. Our results demonstrate a method for high-frequency conversion that could be readily integrated into silicon-based quantum computing systems and be applied to other semiconductors.

\end{abstract}

\maketitle

\section{Introduction}

Quantum dots (QDs) show great promise as a scalable platform for quantum computation~\cite{loss1998quantum} following the recent demonstrations of universal control of 4- and 6-qubit systems~\cite{hendrickx2021four, philips2022universal} and the scaling to 4x4 QD arrays~\cite{borsoi202216}. Quantum dots in silicon as a host for spin qubits are particularly attractive due to the demonstrated high-fidelity control~\cite{xue2022quantum,noiri2022fast,mills2022two},
industrial manufacturability~\cite{Maurand2016, zwerver2022qubits}, co-integration with classical cryogenic electronics~\cite{Charbon2016,xue2021cmos,ruffino2022cryo} and the availability of detailed proposals for large-scale integration~\cite{li2018crossbar, veldhorst2017silicon, vandersypen2017interfacing, gonzalez2021scaling, boter2022spiderweb, crawford2022compilation, fogarty2022silicon}. 

As QDs become an established platform for qubit implementation, novel ways of operating such devices are beginning to emerge, providing an opportunity to integrate on the same embodiment additional functionalities. In particular, by exploiting the non-linear high-frequency admittance, QDs can be utilised as compact charge sensors~\cite{oakes2022fast,Niegemann2022}, fast local thermometers~\cite{Ahmed2018, chawner2021nongalvanic} or as parametric amplifiers~\cite{Cochrane2022} for quantum-limited amplification. Here, we present a new functional implementation: a quantum dot-based radio-frequency multiplier.

A frequency multiplier typically consists of a non-linear circuit element which distorts a sinusoidal signal, generating multiple harmonics. A band-pass filter is then used to select a particular harmonic. Commercially, this can be achieved by utilising, for example, semiconductor-based diodes or amplifiers but these dissipate energy reducing their power conversion performance. Alternatively, microelectromechanical resonators can be utilised for multiplication with enhanced performance due to their low-loss nature~\cite{Nguyen2007}.   

In this manuscript, we propose and demonstrate a QD-based radio-frequency multiplier in silicon. We demonstrate frequency multiplication on a single device in two complementary configurations: a single QD couple to a reservoir and a double QD. Due to the non-dissipative nature of the device impedance under elastic tunnelling conditions, the devices show near-ideal power conversion and phase-noise properties for a multiplication factor of up to 10. Such devices could be utilised to alleviate the challenges associated with the delivery of high-frequency signals to quantum processors from room temperature down to cryogenic stages in dilution refrigerators while reducing the cost and complexity of the room temperature electronics.


\section{QD frequency multiplication:}

\begin{figure}[ht]
    \centering
    \includegraphics[width=1\linewidth]{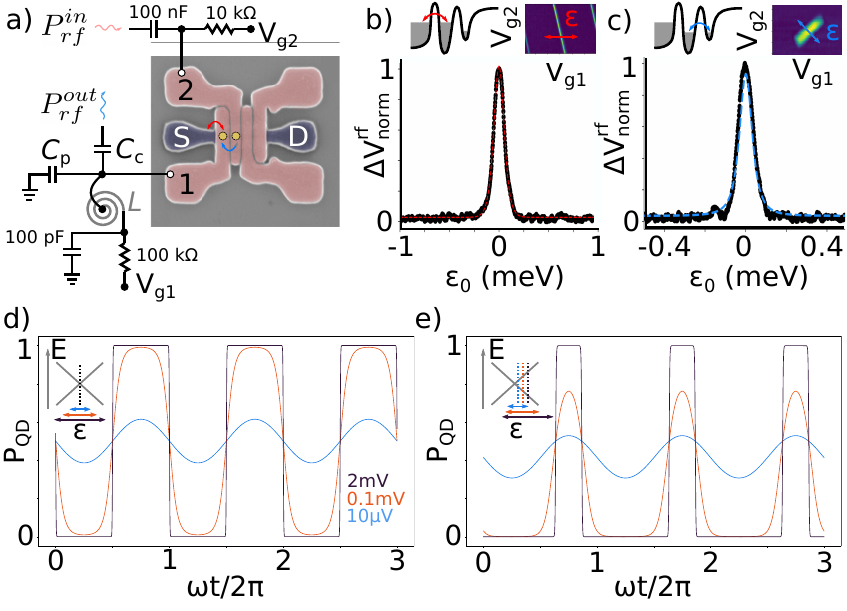}
    \caption{Device and experimental setup. a) False-coloured scanning electron micrograph of a silicon nanowire transistor with wraparound gates, under which QDs form (yellow circles). An rf signal is applied to gate 2 via a fast line. The signal is measured in transmission from gate 1 via a lumped element $LC$ resonator, which acts as a narrow band-pass filter at $f_\text{r}$. Normalised change in rf voltage due to electron tunnelling across a dot-to-reservoir (b) and inter-dot-charge-transition (c) probed at $f_\text{r}$.The insets in both figures schematically represent the tunnelling processes and show the charge stability maps with the relevant transitions and how detuning $\varepsilon$ is defined. Simulations of electron occupation probability of a QD for zero (d) and non-zero (e) detuning $\varepsilon_0$ for different driving amplitudes $\delta \varepsilon$. By strongly driving the system, a non-linear response occurs, tending toward a square, where the duty cycle can be tuned by adjusting $\frac{\varepsilon_0}{\delta \varepsilon}$, which in panel (e) is set to 75\%.}
    \label{fig1}
\end{figure}

The device used to demonstrate frequency multiplication is the same as in ref.~\cite{oakes2022fast}. It consists of a fully-depleted silicon nanowire transistor with four wraparound gates in series, as shown in Fig.~\ref{fig1}a. Quantum dots form under the gates at cryogenic temperatures when biased close to the threshold voltage~\cite{Ibberson2018}. In this demonstration, we utilise voltages on gate 1 ($V_\text{G1}$) and 2 ($V_\text{G2}$) to form either a QD coupled to a reservoir or a double QD. Further, we apply a radio-frequency (rf) signal via gate 2 and then measure the transmission through the QD system at the output of an $LC$ lumped-element resonator connected to gate 1, which acts as a high-quality band-pass filter around its natural frequency, $f_\text{r}=789$~MHz. Changes in transmitted voltage through the setup are directly proportional to the device's admittance $Y$ and since in QD systems, it is a highly nonlinear function of the gate voltage~\cite{Petersson2010,Ciccarelli_2011,Chorley2012,Mizuta2017,Esterli2019, Maman2020}, they can be utilised for frequency multiplication when driven beyond its linear regime, as we shall see. 

We now explain the non-linearity of QD systems in detail. The non-linear admittance arises from the dependence of the gate charge ($Q_\text{g}$) with gate voltage. More particularly, for a QD coupled to a reservoir in the fast relaxation regime, $f\ll\Gamma_\text{R}$, where $\Gamma_\text{R}$ is the charge tunnelling rate, the gate charge reads

\begin{equation}
    Q_\text{g}=e\alpha_{21}P(t) = e\alpha_{21} f(\varepsilon(t)) = \frac{e\alpha_{21}}{\exp[\varepsilon(t)/k_\text{B}T] + 1}
    \label{eq:P_DTR_approx}
\end{equation}

\noindent where $P$ is the electron occupation probability, $f$ is the Fermi-Dirac probability distribution, $k_\text{B}$ the Boltzmann constant, $e$ the charge of an electron and $T$ the system temperature. Here  $\varepsilon(t)=-e\alpha_\text{i1}(V_\text{Gi}-V_\text{Gi}^0)$ is the detuning between the QD electrochemical level and the Fermi level at the reservoir, $V_\text{Gi}^0$ is the voltage at which these two levels are aligned and $\alpha_{ij}$ is the $i^{th}$ gate lever arm to dot $j$~\cite{Ahmed2018} for $i,j=1,2$. In the case of a double quantum dot (DQD), in the adiabatic limit, the gate charge is

\begin{equation}
    Q_\text{g}=e\alpha'P(t) = \frac{e\alpha'}{2}\left[1 + \frac{\varepsilon(t)}{\sqrt{\varepsilon(t)^2 +\Delta^2}}\right] = \frac{e\alpha'}{2}\left[1 + \Pi(t)\right]
    \label{eq:P_ICT}
\end{equation}

\noindent where in this case $\varepsilon(t)$ refers to the energy detuning between QDs, $\alpha'=\alpha_{22}-\alpha_{21}$ to the ICT lever arm, and $\Delta$ is the tunnel coupling~\cite{Mizuta2017}. When subject to the oscillatory detuning [$\varepsilon(t) = \varepsilon_0 + \delta \varepsilon \cos{\omega t}$] due to the rf excitation, an oscillatory gate current flows through the system proportional to its admittance

\begin{equation}
    I_\text{g}=e\alpha\frac{dP(t)}{dt}= Y V_\text{g}. 
    \label{eq:GateI}
\end{equation}

Changes in device admittance occur either due to dot-to-reservoir (DTR) charge transition, as shown in Fig.~\ref{fig1}b or inter-dot-charge transitions (ICTs), depicted in Fig.~\ref{fig1}c with their corresponding fits using the calculated admittance. If the rf excitation is small ($\delta\varepsilon \ll k_\text{B}T, \Delta$), the DTR or ICT respond linearly to the input voltage. In this regime, the charge oscillates at the rf frequency, with a phase shift dependent on the magnitude of the Sisyphus resistance~\cite{Esterli2019}, see blue traces in Fig.~\ref{fig1}(d,e). When driving the systems beyond their linear regime, the charge distribution gets distorted as the QD is either filled or emptied. As long as $\delta \varepsilon$ remains smaller than the charging energy, the system is required to obey $0 \leq P \leq 1$ at all times. Eventually, in the large-signal regime ($\delta\varepsilon \gg k_\text{B}T, \Delta$), $P(t)$ acquires a square wave-like response containing high-frequency harmonics of the original signal, see Fig.~\ref{fig1}d,e and Appendix~\ref{app:Square}. Comparing Eq.~\ref{eq:P_DTR_approx} and Eq.~\ref{eq:P_ICT}, we note how the DTR depends exponentially on $\varepsilon$, while $\Pi(\varepsilon)$ is a polynomial function. Therefore, we expect the DTR to show a higher degree of non-linearity than the ICT in the intermediate signal regime, as we shall see.

Following the derivation in Appendix \ref{app:Y}, we can then express the rf voltage output at the $N$-th harmonic, $|\Delta V^\text{rf, out}_N|$, in terms of the $N$-th harmonic of the gate current and hence that of the charge occupation probability,
\begin{align}
    |\Delta V^\text{rf, out}_N| \propto \left|\int_0^{\frac{2 \pi}{N \omega}} e^{i N \omega t} P(t) dt \right| \equiv P_N.
    \label{eq:V_from_P}
\end{align}

For a DTR with tunnel rate $\Gamma_R$, a simple analytical solution can be found for the output voltage at the $N$-th harmonic, 
\begin{align}
    |\Delta V^\text{rf, out}_N| \propto \left|\frac{\Gamma_\text{R}}{\Gamma_\text{R} + i N \omega} ~ f_N  \right|,
    \label{eq:V_DTR_true}
\end{align}
\noindent
where we define $f_N$ as
\begin{equation*}
    f_N = \frac{\omega}{2 \pi} \int_0^{\frac{2 \pi}{\omega}} e^{i N \omega t} f(\varepsilon_0 + \delta \varepsilon \cos{\omega t}).
    \label{eq:Fcomp}
\end{equation*}

For the ICT, there is no simple general analytical solution for the output voltage, equivalent to Eq.~\eqref{eq:V_from_P} (see Appendix~\ref{app:Square}). However, using a semiclassical approach, we derive a simple expression for the driven probability distribution. In the regime of fast relaxation, i.e. $\Gamma_\text{P} \gg \omega$, where $\Gamma_\text{P}$ is the coupling rate to the phonon bath, absorption and emission mix the ground and excited states, driving them to thermal equilibrium faster than the basis rotation caused by the rf perturbation~\cite{Esterli2019}. If we define $n_0 = b\left(\sqrt{\Delta^2 + \varepsilon_0^2}\right)$, with $b(E) = \frac{1}{e^{E/k_\text{B}T} - 1}$ being the Bose-Einstein distribution, we find a simple expression for the driven probability distribution
\begin{equation}
    P(t) = \frac{1}{2} + \frac{1}{2 n_0 + 1}\Pi(t).
    \label{eq:P_ICT_thermal}
\end{equation}
Equation~\eqref{eq:P_ICT_thermal} presents the same large-signal square-wave limit discussed thus far but with a different response to the DTR in the intermediate regime. We note the equation is valid in the limit of $\Delta^2 \gg \hbar \omega \delta \varepsilon$, where the evolution can be considered adiabatic.

Finally, we find that in the large signal regime, ($\delta\varepsilon \gg k_\text{B}T, \Delta$), both systems, QD and DQD, produce an output voltage given by  

\begin{equation}
     |\Delta V^\text{rf, out}_N| \propto \frac{1}{N} \sin{\left[ N~ \text{arccos}\left(\frac{\varepsilon_0}{\delta \varepsilon}\right)\right]}, 
     \label{eq:large_V}
\end{equation}
\noindent
which unifies the large-signal response of the DTR and ICT as it only depends on $\varepsilon_0$ and $\delta \varepsilon$ and not on the physical characteristics of the particular quantum systems. Equation~\eqref{eq:large_V} also highlights that the maximum voltage conversion at the $N$-th harmonic follows a $N^{-1}$ trend, which translates into the maximum $N^{-2}$ trend for the well-known maximum power conversion efficiency of an ideal frequency multiplier. Further, we note that the detuning offset, $\varepsilon_0$, changes the duty cycle $DC$ of the square-like output wave (see Appendix ~\ref{app:Square}) as we vary the probability of the QD being filled or emptied, as shown in Fig.~\ref{fig1}e. Accordingly
\begin{equation}
    DC = \frac{1}{\pi} \text{arccos}\Big(\frac{\varepsilon_0}{\delta \varepsilon}\Big),
    \label{eq:DC}
\end{equation}
\noindent if $\delta \varepsilon \geq |\varepsilon_0|$, while the QD remains trivially empty or full if $\varepsilon(t)$ never changes sign. The offset detuning dependence of $DC$, and thus $\Delta V^\text{rf, out}_N$, enables changing the relative amplitude of the spectral components of the voltage output, as we shall see.

\section{Experimental Results}

\begin{figure*}[ht]
    \centering
    \includegraphics[width=\linewidth]{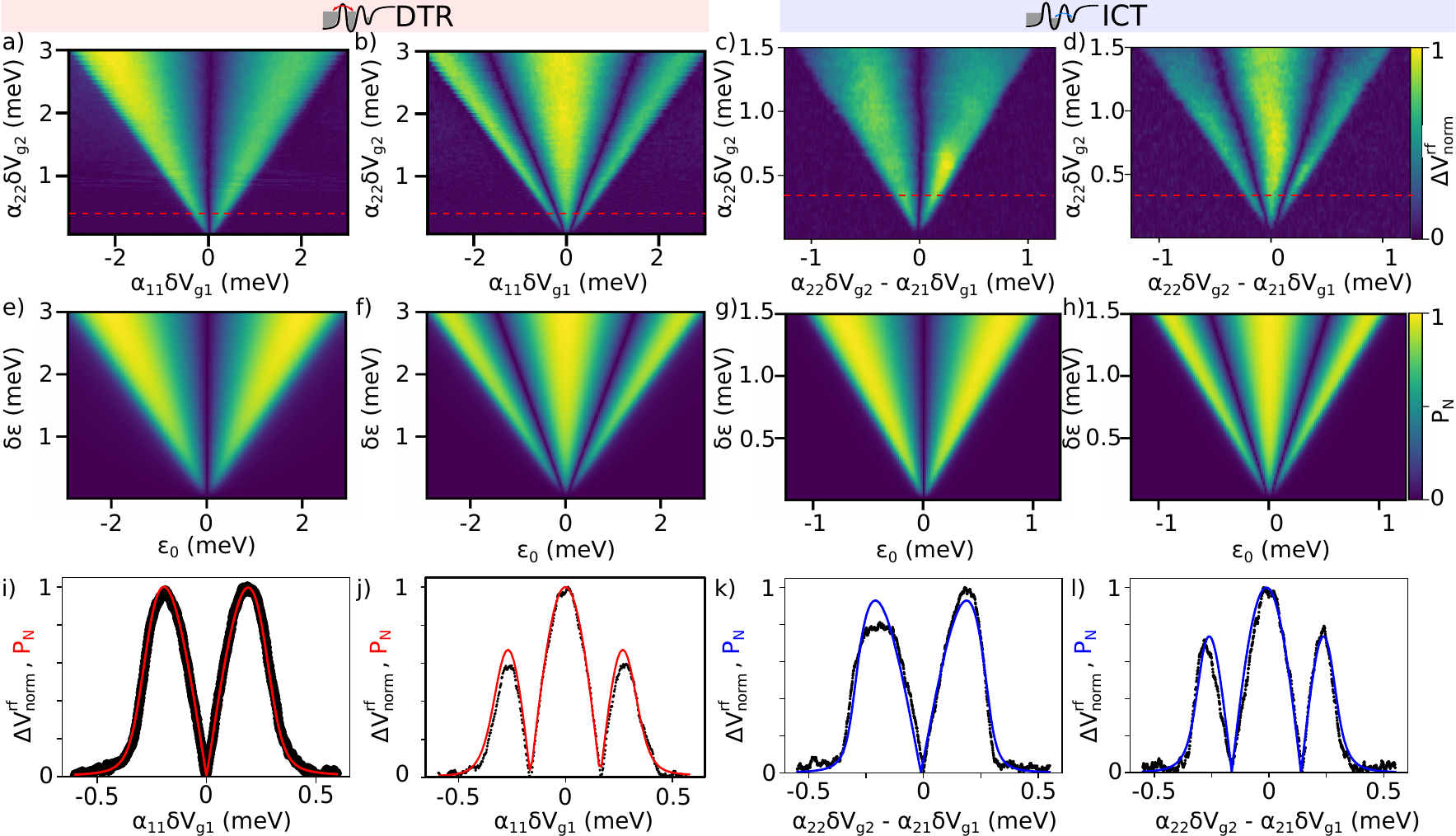}
    \caption{QD-based frequency multiplier line-shapes. Experimentally measured second (a,c) and third (b,d) harmonic line-shape as a function of driving rf signal for the DTR and ICT respectively. e-h) Simulations of the corresponding panels above were carried out to demonstrate experimental agreement with the theory. i-l) Are line cuts at a driving rf amplitude of 0.37~meV peak-to-peak (red dotted line in panels a-d). Data points are in black and fits are solid lines.}
    \label{fig2}
\end{figure*}

Equation~\eqref{eq:V_from_P} shows how large rf perturbations can be used to demonstrate a QD-based frequency multiplier. Due to the fixed resonance frequency of our resonator ($f_\text{r}\sim$789 MHz), which acts as a sharp band-pass filter (see Appendix~\ref{app: band-pass}), we fix the output frequency by setting the local oscillator (LO) used for homodyne mixing at $f_\text{r}$. To probe the various harmonics, we perturbed the system via an rf signal along gate 2 at integer fractions of the resonator's frequency $f=f_\text{r}/N$. In Fig.~\ref{fig2}, we show how the second and third harmonics evolve as a function of driving $\delta \epsilon$ and detuning offset $\varepsilon_0$ for a DTR and an ICT. We observe that all harmonics are symmetric along zero detuning (up to a phase of $(N+1) \pi$) and are characterised by $N$ lobes in the magnitude response signal. For the even harmonics, the maximum voltage transferred occurs at $\varepsilon_0=\pm \sin{\left(\frac{\pi}{N}\right)} \delta \varepsilon$ and is zero at $\varepsilon_0=0$ (see Appendix~\ref{app:alpha_rf}). In contrast, for the odd harmonics, the signal is maximum at $\varepsilon=0$. As a result, the optimum detuning point to operate frequency multiplication varies for the different harmonics.

To accurately simulate the line-shape response for the various harmonics for both DTRs and ICTs, we developed a Lindblad framework. The method entails iterating the Lindblad Master Equation~\cite{manzano_short_2020}, defined as
\begin{equation}
\hbar \frac{d}{dt} \rho = -i [H(t), \rho] + \sum_{i = +,-} \hbar \Gamma_i \mathcal{D}[L_i]
\label{eq:LME_main}
\end{equation}
\noindent with
\begin{equation}
    \mathcal{D}[L_i] = L_i^\dagger \rho L_i - \frac{1}{2} \left\{ L_i^\dagger L_i, \rho \right\}
\end{equation}
\noindent to numerically obtain the QD occupation probability $P(t) = \rho_{1, 1}(t)$. Combined with Eq.~\ref{eq:V_from_P}, it allows us to simulate the rf reflectometry lineshapes without making use of the approximations in Eqs.~\eqref{eq:P_DTR_approx} and \eqref{eq:P_ICT}. A mathematical derivation of the \textit{jump} operators $L_i$ and their relative relaxation rates $\Gamma_i$ for the DTR and the ICT can be found in Appendix~\ref{app:LB}. The gate lever arms ($\alpha_{11}=0.59\pm0.02$ and $\alpha_{\text{ICT}}=0.26\pm0.01$) and electron temperature ($T_e=140\pm 3$~mK) were extracted experimentally in the small signal regime (see Appendices~\ref{app:alpha}-~\ref{app:Te}), leaving the reservoir-dot relaxation rate $\Gamma_\text{R}$ and the double QD tunnel coupling $\Delta$ as fitting parameters. 
Because of the different capacitive coupling between gates at high frequency, we estimate from the data the high-frequency lever arm of G2 on the QD along the DTR, with a value of $\alpha^\text{rf}_\text{21} \approx 0.35$. 
With these values, we use lineshapes at constant rf power (Fig.~\ref{fig2}(i-l)) to fit $\Gamma_\text{R}= 2.9 \pm 0.6$ GHz and the ICT tunnel coupling $\Delta=35 \pm 9$~$\mu$eV. 

Comparing Fig.~\ref{fig2}(i-l), we notice the stark similarity of the lineshapes of the DTR and ICT in the large-signal regime ($\delta \varepsilon = 0.37$ meV), which can be attributed to the same square-wave-like behaviour of Eq.~(\ref{eq:P_DTR_approx}) and Eq.~(\ref{eq:P_ICT}). We see, however, slightly more pronounced sidebands in the ICT at the 3$^{\text{rd}}$ harmonic when compared to the DTR. We attribute this to the lifetime-broadening of the DTR (see Fig.~\ref{fig:broadening} in Appendix~\ref{app:Te}), whose model (red line) shows excellent agreement with the experimental data. Moreover, comparing the data with the fits (blue lines) for the ICT, we observe an asymmetry with respect to $\varepsilon_0=0$, which is not present in the theory.
We attribute this effect to not having swept the gate voltages perfectly along the detuning axis due to the difference in lever arms between G1 and G2.
\begin{figure}[ht]
    \centering
    \includegraphics[width=1\linewidth]{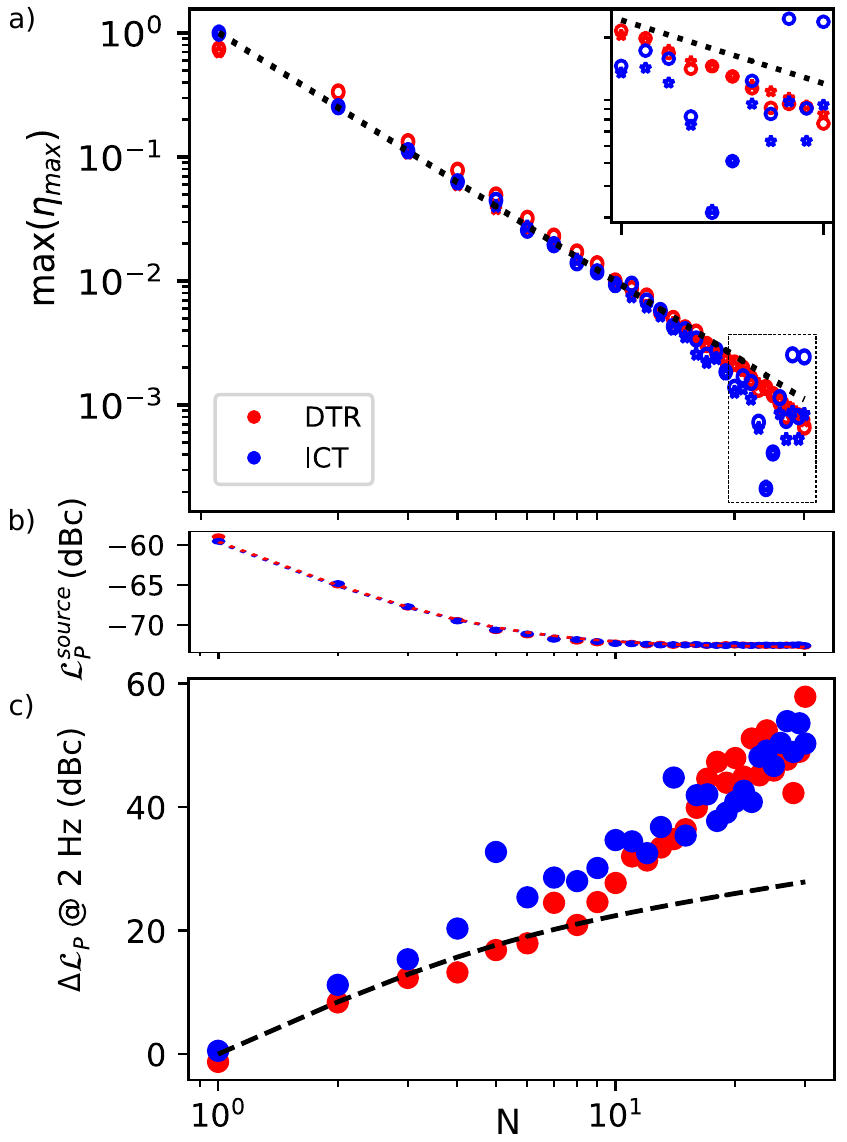}
    \caption{Performance of a QD-based frequency multiplier. a) Maximum power conversion efficiency max($\eta_{\text{max}}$) compared to the first harmonic for the DTR (red) and the ICT (blue) for increasing harmonic $N$. Experimentally measured values are plotted as circles and the fitted saturation points as stars. The first 20 harmonics closely follow an $N^{-2}$ dependence (black dashed line). The inset zooms into the region between $N=20$ and $N=30$. b) Phase noise of the rf source at the relevant rf powers used for the DTR and ICT. The phase noise measured follows Leeson's equation (dotted line). c) Difference in phase noise caused by the QD-based frequency multiplier, which increases as a function of multiplication $N$. The black dotted line represents the ideal phase noise for a fully saturated signal (see Appendix~\ref{app:phase_noise}).}
    \label{fig4}
\end{figure}

We now move on to assess the performance of the QD-based frequency multiplier. First, we investigate the harmonic power conversion efficiency by rf voltage output maps equivalent to those shown in Fig.~\ref{fig2} up to $f_\text{r}$/50 (Appendices~\ref{app:DTR_50}-~\ref{app:ICT_50}).
We quantify this efficiency by defining the figure of merit
\begin{equation}
    \eta_{\text{max}}(\delta \varepsilon) = \frac{\text{max}_{\varepsilon_0}\left(|V^{rf, out}_N|^2\right)}{\text{max}_{\varepsilon_0}\left(|V^{rf, out}_1|^2\right)}
    \label{eq:eta}
\end{equation}
\noindent
where $V^{rf, out}_N$ indicates the measured output voltage at the $N^{\text{th}}$ harmonic. We chose this metric as it accounts for losses along the rf lines. We take the maximum signal for a given rf input power, as the detuning $\varepsilon_0$ for optimal conversion, which depends on the specific harmonic of interest (Fig.~\ref{fig2}). At $f_\text{r}$/50, a weak signal is still discernible for the DTR transitions but not for the ICTs, which we detect up to $\sim f_\text{r}$/30.
We then extract the power conversion efficiency for the different harmonics at a given input power (Appendix~\ref{app:power_conversion}). In Fig.~\ref{fig4}a, we plot the maximum power conversion efficiency $\eta_{\text{max}}$ compared to the first harmonic for the DTR and the ICT. The maximum power conversion efficiency follows the expected $N^{-2}$ dependence for a square wave (black dotted line) until $\sim f_\text{r}$/20, above which max($\eta_{\text{max}}$) begins to drop below the $N^{-2}$ trend, as highlighted by the insert. 
We attribute lower power conversion efficiency observed at higher harmonics to the signal not being fully saturated at the maximum rf power used to probe the device. The ICT diverges from the $N^{-2}$ dependence more drastically than the DTR due to a combination of factors. Firstly, being unable to drive the system as hard before excited states and multiple charge transitions start interfering (see Appendix~\ref{app:n**2}). Secondly, there is a decrease in signal due to the less-pronounced non-linearity for an ICT compared to a DTR transition (see Appendices~\ref{app:DTR_50}-\ref{app:ICT_50}).

Another important metric for frequency multiplication is to assess the increase in phase noise $\mathcal{L}_p$ at higher harmonics. To calibrate our system, we measure the phase noise of the rf source, $\mathcal{L}_p^\text{source}$, (Fig.~\ref{fig4}b) at the frequencies used to drive the DTR and ICT. The decrease in phase noise as we decrease the excitation frequency ($f=f_\text{r}/N$) follows the empirically established Leeson's equation~\cite{lee2000oscillator} (red dotted line). 

In Fig~\ref{fig4}c, the data shows the phase noise contribution of the QD-based frequency multiplier, $\Delta\mathcal{L}_p$, obtained by subtracting the rf source phase noise from the total phase noise measured. The harmonics produced by the driven QD system have an increasing phase noise as a function of the multiplication factor.
Up to $N\sim$10, the phase noise of our system is ideal (black dotted line), as defined in Appendix~\ref{app:phase_noise}, above which, it diverges. The exact reason is unknown but we speculate that it is due to the output signal not being fully saturated. In this regime, additional phase noise may arise due to charge noise since the system is more sensitive to charge probability variations as the occupation probability is not in a square-wave-like regime.

\begin{figure}[ht]
    \centering
    \includegraphics[width=1\linewidth]{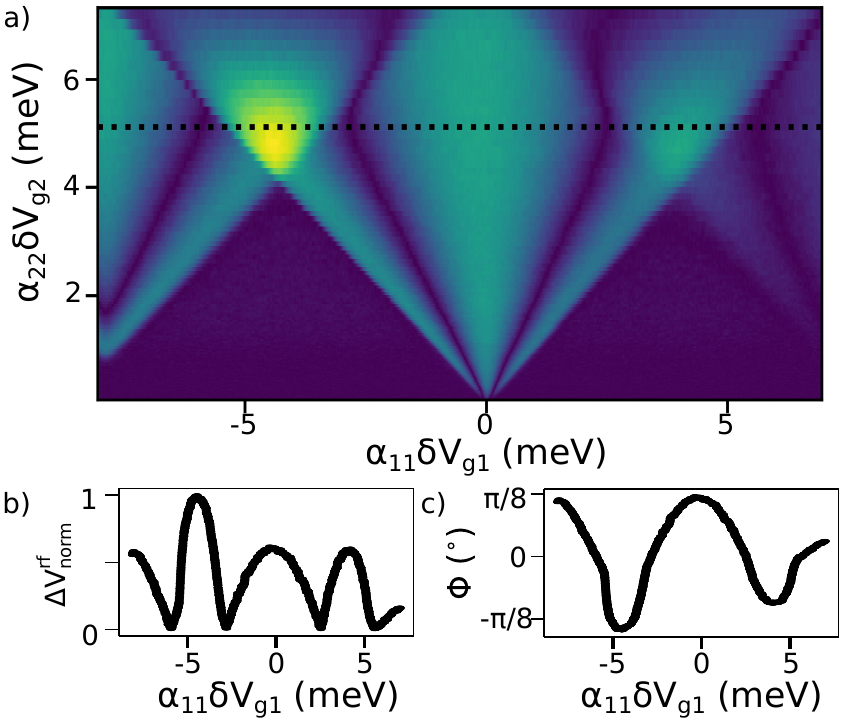}
    \caption{Interference between multiple electron transitions. a) Harmonic line shape for $f_\text{r}/3$ of multiple DTR lines as a function of driving rf signal. By driving the system hard enough, the DTR lines interfere, resulting in regions of constructive interference with max($\eta_{\text{max}}$) exceeding the expected N$^{-2}$ dependence, as highlighted by the line-cuts in b) magnitude and c) phase.}
    \label{fig5}
\end{figure}

Finally, we examine the performance of the multiplier when multiple charge transitions begin to merge above a given driving power, ($P>1$). In Fig.~\ref{fig5}a, we explore $N=3$. We observe that at an excitation amplitude $\sim 4$~meV, neighbouring DTR lines begin to interfere; as two electrons are being loaded and unloaded in phase from the QD.
Due to the $N$-1 nodes present in the phase of the $N$-th harmonic, constructive and destructive interference can occur, which can give rise to an increase in signal proportional to the number of tunnelling electrons involved in the process, resulting in max($\eta_{\text{max}}$) exceeding the expected $N^{-2}$ dependence. Within this regime, the two-level system framework developed in Appendices~\ref{app:Square} and~\ref{app:Y} is no longer valid. Therefore, the data in this regime was omitted from the max($\eta_{\text{max}}$) and $\mathcal{L}_p$ discussion. We also observe the same interference behaviour for the ICT, with the added phenomenon of having excited states appearing at certain driving rf powers, as outlined in Appendix~\ref{app:n**2}. Although harder to interpret, these regions of interference allow higher harmonics to be measured. We detected rf output signals up to $f_\text{r}/75$ (Appendix~\ref{app:n**2}).

\section{Conclusions and outlook} \label{sec: conclusions}

We have introduced the idea of a quantum-dot based frequency multiplier and demonstrated its experimental implementation via two methods: a QD coupled to a reservoir, and a double QD. We find near-ideal performance in terms of power conversion and phase noise up to $N\approx 10$. The device may facilitate generating on chip the high-frequency signal needed for qubit control ($4-40$~GHz)~\cite{Zhao2019,Veldhorst2014} and readout (0.1-2~GHz)~\cite{Vigneau2022} by operating it in conjunction with a lower frequency generator. Such an approach may help preserve signal integrity from room temperature down to the mixing chamber of a dilution refrigerator and/or may facilitate the development of an integrated cryogenic frequency generator. Furthermore, if combined with frequency mixing~\cite{Colless2013,Gonzalez-Zalba2015,Ares2016} and parametric amplification~\cite{Cochrane2022}, one may envision an all QD-based rf reflectometry setup including frequency generation, amplification and mixing. 

With regard to the performance, we expect near-ideal properties as long as the electron tunnelling process on which the operation of the multiplier is based remains elastic. This condition sets upper bounds for the maximum input frequency to $\omega\ll\Gamma_\text{R}$ in the DTR case, and
$\Delta \gg \sqrt{\hbar \omega \delta\varepsilon}, k_B T$
for the ICT. The device remains operational at higher frequencies but additional dissipation may deteriorate its performance.
Finally, we note that the large-signal regime explored here can be utilised for excited state spectroscopy of (D)QD systems. 

\section{Acknowledgements}

This research was supported by European Union’s Horizon 2020 research and innovation programme under grant agreement no.\ 951852 (QLSI), and by the UK's Engineering and Physical Sciences Research Council (EPSRC) via the Cambridge NanoDTC (EP/L015978/1), QUES2T (EP/N015118/1), the Hub in Quantum Computing and Simulation (EP/T001062/1) and Innovate UK [10000965].
 M.F.G.Z. acknowledges a UKRI Future Leaders Fellowship [MR/V023284/1].

\appendix

\section{Occupation in the large-signal regime}
\label{app:Square}
\begin{figure*}
    \centering
    \includegraphics[width=\linewidth]{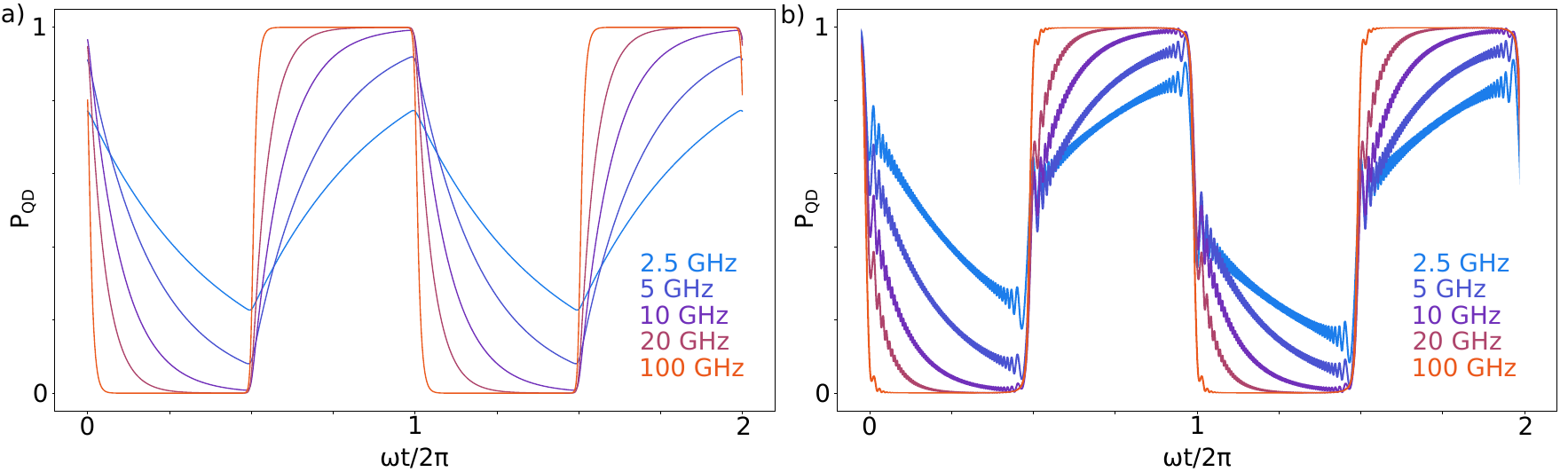}
    \caption{Lindblad Simulations (see Appendix~\ref{app:LB}) of the steady-state QD occupation for a DTR (a) and an ICT (b) for $\varepsilon_0 = 0$ and $\delta \varepsilon = 1$ meV at an rf frequency of $\frac{\omega}{2 \pi} =$1 GHz at $140$ mK. The different curves indicate $\frac{\Gamma_R}{2 \pi}$ for the DTR and $\frac{\Gamma_P}{2 \pi}$ for the ICT. (a) Electronic occupation of the QD evolves from a square wave for $\Gamma_r \gg \omega$ to a low-pass filtered square wave for $\Gamma_R \sim \omega$, as predicted by Eq.~(\ref{eq:P_DTR_sq}). (b) Electronic occupation of one of the two QDs across an ICT (i.e. $\rho_{0,0}$(t)) The tunnel coupling has been set to $\Delta = 35 \mu$eV as fitted from the experimental data. Comparing panels (a) and (b) shows the qualitative similarity for the two systems as $\Gamma$ approaches $\omega$. However, additional high-frequency oscillations at the Rabi frequency $\frac{E_+ - E_-}{\hbar} \gg \omega$ appear as phonons relax the system into the ground state after Landau-Zener transitions. We note that in these simulations dephasing has been intentionally ignored to emphasise this effect, which cannot be captured by semiclassical models.}
    \label{fig:lowpass}
\end{figure*}

The naive solution to the (D)QD occupation in the large signal regime would be to assume a square-wave-like behaviour, where the (D)QD is periodically emptied and filled if the rf amplitude is much larger than the detuning. This ansatz, however, fails due to the finite relaxation times of the respective systems.
For the DTR, we can find an analytical expression for the large-signal regime through the Master Equation in Eq.~(\ref{eq:Master_semicl}), whose solution at the steady-state is
\begin{equation}
    P(t) = \Gamma_\text{R} e^{-\Gamma_\text{R}t} \int_{-\infty}^{t}e^{\Gamma_\text{R}\xi} f(\varepsilon(\xi)) d\xi
    \label{eq:P_DTR_integrals}
\end{equation}
\noindent
which can be expressed as the Fourier series
\begin{equation}
    P(t) = f(\varepsilon_0) + \frac{1}{2} \sum_{n=1}^{\infty} \left(\frac{\Gamma_\text{R}}{\Gamma_\text{R} + i N \omega} ~ f_N ~ e^{ i N \omega t} + c.c. \right)
\label{eq:P_DTR_sq}
\end{equation}
\noindent
where
\begin{equation*}
    f_N = \frac{\omega}{2 \pi} \int_0^{\frac{2 \pi}{\omega}} e^{i N \omega t} f(\varepsilon_0 + \delta \varepsilon \cos{\omega t})
\end{equation*}
is the $N$-th Fourier component of the perturbed Fermi-Dirac.
In the limit of fast relaxation ($\Gamma_\text{R} \gg \omega$), Eq.~\ref{eq:P_DTR_sq} simplifies to $P(t) = f(\varepsilon(t))$, which indeed tends to a square wave for large driving. For $\Gamma_\text{R} \sim \omega$ the coupling to the reservoir behaves as a low pass filter for the square wave (see Fig.~\ref{fig:lowpass}(a)). 
In our particular experimental setup we have a fixed resonator, thus $N \omega = \omega_r$ is constant. Therefore, the prefactor 

\begin{equation*}
    \frac{\Gamma}{\Gamma + i N \omega} = \frac{\Gamma}{\Gamma + i \omega_r},
\end{equation*}
\noindent
is just a global constant for all harmonics and all powers, thus all the dynamics is described by $f_N$, which is square-wave-like.

For the ICT, we can consider a semiclassical solution in which the system evolves while remaining in the ground state. This represents the behaviour of the ICT if the tunnel coupling is large enough to neglect Landau-Zener transitions (i.e. $\Delta^2 \gg \hbar \omega \delta \varepsilon$)~\cite{SHEVCHENKO2010} and relaxation events (i.e. $ \Delta \gg \hbar \Gamma_\text{P} \left[ 2 b(E) + 1 \right]$).
In this case, we can simply write:

\begin{equation}
    P(t) = \frac{1}{2}\left[1+\frac{\varepsilon(t)}{\sqrt{\varepsilon(t)^2 +\Delta^2}}\right] = \frac{1}{2}\left[1+\Pi(t)\right]
    \label{eq:FD_FT}
\end{equation}
\noindent
where $P(t)$ is the probability of the electron occupying the QD under the input gate. In the large-signal regime (i.e. $\delta \varepsilon \gtrapprox \Delta$), it is clear that $\Pi(t)$ tends to a square wave as the charge shifts from one QD to the other.

When considering the full model from the LME, we find the possible presence of Landau-Zener transitions when $\delta \varepsilon \approx \frac{\Delta^2}{\hbar \omega}$, which produce oscillations at the Rabi frequency $\Omega_\text{R} = (E_+ - E_-)/\hbar$. The result of these two effects produces a $P(t)$ which is formally no longer a square wave.
However, we can gain insight into the effect of phonon relaxation in the system by considering the semiclassical approximation of the LME. In the basis where $H(t)$ is instantaneously diagonal (i.e. the \textit{energy basis}), this reads
\begin{equation}
    \dot{P}_\text{EB} + \Gamma_\text{P} (2 b(E(t)) + 1) P_\text{EB} =  \Gamma_\text{P} (b(E(t)) + 1)
    \label{eq:Master_ICT}
\end{equation}
whose formal solution is
\begin{equation}
    P_\text{EB}(t) = \Gamma_\text{P} e^{-\Gamma_\text{tot}(t)} \int_{-\infty}^{t}e^{\Gamma_\text{tot}(\xi)} (b(E(\xi)) + 1) d\xi
    \label{eq:P_ICT_integrals}
\end{equation}
\noindent
where we defined 
\begin{equation}
    \Gamma_\text{tot}(t) = \Gamma_\text{P} \int (2 b(E(t)) + 1) dt
\end{equation}

Projecting back into the QD basis, we can write 
\begin{align}
    &P(t) \approx \frac{1}{2}\biggr[1+\frac{1}{2 n_0 +1} \Pi(t) -\\
    &- \sum_{N=1}^{\infty} \left(  \left( \left\{ \frac{\Gamma_\text{P} (n_0+1)b_q}{\Gamma_\text{P} (2n_0+1) + i q \omega} \right\} * \left\{ \Pi_{q}  \right\} \right)_N ~ e^{ i N \omega t} + c.c. \right)\biggr]\nonumber
    \label{eq:P_ICT_sq}
\end{align}


\noindent
where $\left(\{a_q\}*\{b_q\}\right)_N$ indicates the $N$-th Fourier component of the convolution of the two Fourier series. 
Comparing the formal similarity Eqs.~(\ref{eq:P_DTR_integrals}) and (\ref{eq:P_ICT_integrals}), we can infer that the two systems will have comparable responses to increasing relaxation rate. More thoroughly, it is possible to put a boundary on the effect of relaxation by considering that $|b_q| \leq \frac{b(2 \Delta)}{q}$. In particular, in the large-signal $\delta \varepsilon \gg \Delta$, $b(E)$ is meaningfully different from zero only on the points in the cycle where $E(t) \lessapprox k_\text{B}T$. Therefore, the effect of the convolution is to add a finite slew rate to the square-wave-like response of $\Pi(t)$. Therefore, we ought to be able to neglect the effects of $b_{q>1}$, recovering the exact equivalence of the ICT with the DTR. 
This is highlighted in Fig.~\ref{fig:lowpass}(b), which shows how the behaviour of the ICT remains qualitatively similar to that of the DTR when including relaxation and for varying $\Gamma_\text{P}$. Moreover, we can see the fast oscillations at the Rabi frequency, which cannot be captured by the semiclassical nature of the Master Equation in Eq.~\ref{eq:Master_ICT}.


Following Eq.~(\ref{eq:FD_FT}), in the large-signal regime, we expect the main difference between the ICT and the DTR to be, apart from global constants, the substitution of $f(\varepsilon)$ with $\Pi(\varepsilon)$. The fact that the latter is a \textit{polynomial} function of energy, explains the lower degree of non-linearity observed for the ICT when compared with the exponential behaviour of the Fermi-Dirac in the DTR.

Moreover, for both systems, the square-wave behaviour can be written as the Heaviside theta $\theta(\varepsilon(t))$. Therefore, if $\delta \varepsilon \geq |\varepsilon_0|$, it is trivial to predict the duty cycle as 

\begin{equation}
    DC = \frac{1}{\pi} \arccos{\left(\frac{\varepsilon_0}{\delta \varepsilon}\right)}
    \label{eq:DC_A}
\end{equation}

\section{Reflectometry Lineshapes}
\label{app:Y}

Rf reflectometry measures charge transitions via changes in the reflection coefficient of a quantum system. In this work, we measured the \textit{transmission} line-shapes from G2 to G1, as detailed in Fig.~\ref*{fig1}. We can write the transmission coefficient as 

\begin{equation}
    T = \frac{2}{1 + Y Z_0} \approx 2( 1 - Y Z_0)
    \label{eq:Transmission}
\end{equation}
\noindent
where $Z_0 = 50$ $\Omega$ is the characteristic impedance of the line. $Y$ is the effective admittance of the quantum system, where we can assume $|Y|^{-1} \sim \frac{e^2}{h} \gg Z_0$. 
The quantum admittance can be modelled as the charge-dependent equivalent to the Quantum Capacitance and Sisyphus resistance. For a particular harmonic $N$, the quantum admittance $Y_N$ is defined as 

\begin{equation}
    Y_N = \frac{I_N}{\langle V \rangle} = 2 \frac{\alpha^{rf} e^2}{\delta \varepsilon} I_N
    \label{eq:Y}
\end{equation}
\noindent
where $I_N = i N \omega P_N$ is the $N$-th Fourier Component of the current through the device. Therefore, we can calculate the output of the rf measurement as 
\begin{align}
    &|\Delta V^{rf, out}_N| \approx |Y_N Z_0 V^{rf, in}| \propto \\
    &\propto \left|\int_0^{\frac{2 \pi}{N \omega}} e^{i N \omega t} P(t) dt \right|
\end{align}
\noindent
Therefore, we can compute the rf lineshapes of $P(t)$ by either numerically iterating the LME or from the expressions in Appendix \ref*{app:Square}.

We must be aware that the above theory assumes a 0D discrete level with a delta-like density of states. In the case of the DTR, however, the coupling with the reservoir broadens the level. This can be modelled with an effective Lorentzian density $\mathcal{D}(E)$ of states with full-width half-maximum $\Gamma_\text{R}$. This lifetime-broadening becomes important when $\hbar \Gamma_\text{R} \gtrapprox k_\text{B}T$ and can be accounted for considering the convolution of $Y_N(\varepsilon_0)$ for the delta-like level with $\mathcal{D}(\varepsilon_0)$. The DTR of interest in this work is lifetime-broadened, as discussed in Appendix \ref*{app:Te}; therefore, this correction was used to fit the maps in Fig~\ref*{fig:DTR}, with an estimated $\Gamma_\text{R}$ of 2.9$\pm$0.6 GHz.

\section{Frequency band-pass filter:}
\label{app: band-pass}
\begin{figure}[ht]
    \centering
    \includegraphics[width=1\linewidth]{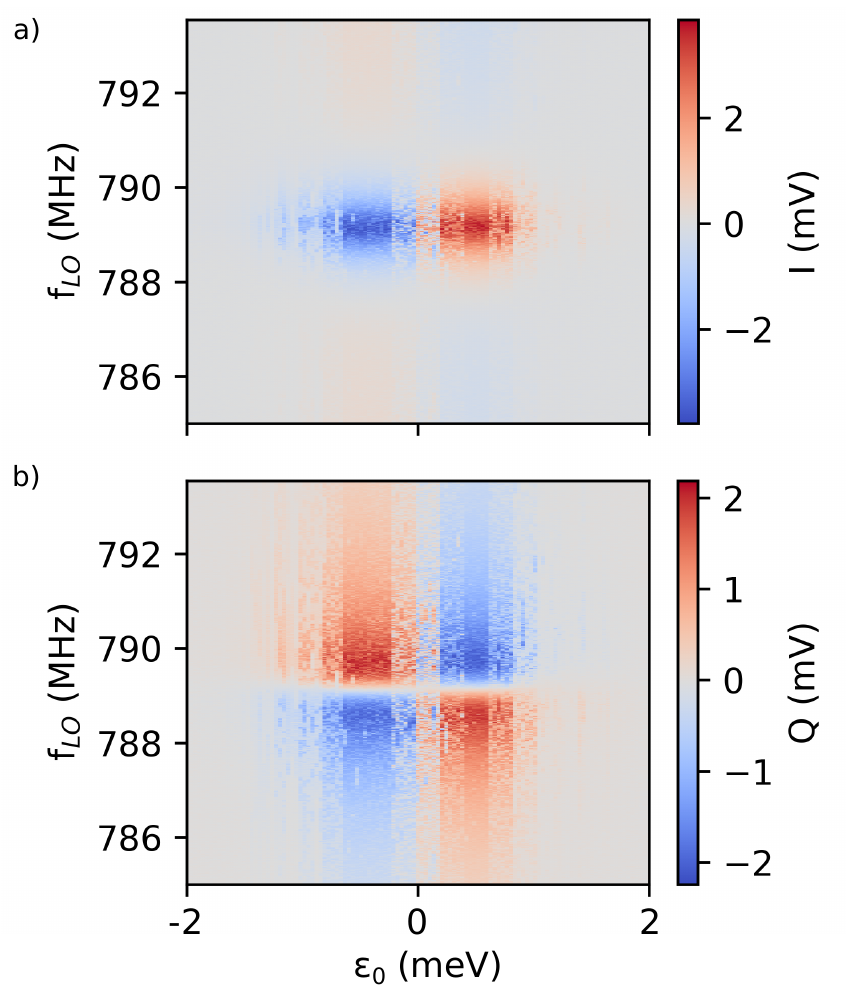}
    \caption{RF signal response in I (a) and Q (b) while probing $\omega_r/2$ at gate 2 and varying the LO frequency $\omega$ on a DTR.}
    \label{fig:freq_bandpass}
\end{figure}

Due to the high-quality factor of the resonator, it acts as a narrow band-pass filter, meaning that even if the device is probed at multiple harmonics, we can only measure at the resonator's frequency. To demonstrate this, we drive gate 2 at $f_\text{r}/2$ while sweeping the frequency of the local oscillator LO of the IQ card around $f_\text{r}$, as shown in Fig.~\ref{fig:freq_bandpass}, illustrating the 5 MHz bandwidth of the resonator, within which we can get a signal. By plotting the in-plane (I) and quadrature (Q) components of the rf signal, we highlight how the line shape is anti-symmetric, resulting in a phase component which can lead to constructive and destructive interference when multiple transitions overlap, as demonstrated in Fig.~\ref{fig5} and highlighted in Appendix~\ref{app:n**2}.

\section{AC Lever arm extraction:}
\label{app:alpha_rf}
From fitting detuning $\varepsilon_0$ vs driving $\delta \varepsilon$ maps, such as those in Fig.~\ref{fig2}, we can experimentally determine the high-frequency lever arms, as well as estimate the power loss of the fast lines. These can differ from those measured in DC through magnetospectroscopy because of different cross-capacitive couplings at high frequencies between the gates, as well as radiative coupling through the resonator via excitations to the first harmonic.
The losses down the fridge can be determined by fitting the ICT power-broadening map, as the main effect of coupling between G1 and G2 is a rigid shift in the total energy of the DQD, which does not affect the reflectometry signal. This result has been obtained by comparing the slope $\varepsilon_0(\delta \varepsilon)$ where the power conversion of the $N$-th harmonic is zero. From Eq.~(\ref*{eq:DC}), the $N-1$ zeros of the $N$-th harmonic should occur for 
\begin{equation}
    \varepsilon_0 = \cos{\left(\pi \frac{m}{N} \right)} ~\delta \varepsilon ~~~~~~ m = 1, ..., N-1
    \label{eq:zeros}
\end{equation}

We also note that the $N$ maxima are predicted to appear for 
\begin{equation}
    \varepsilon_0 = \cos{\left(\pi \left(\frac{m}{N} + \frac{1}{2}\right)\right)} ~\delta \varepsilon ~~~~~~ m = 0, ..., N-1
    \label{eq:maxima}
\end{equation}

\begin{figure}[ht]
    \centering
    \includegraphics[width=1\linewidth]{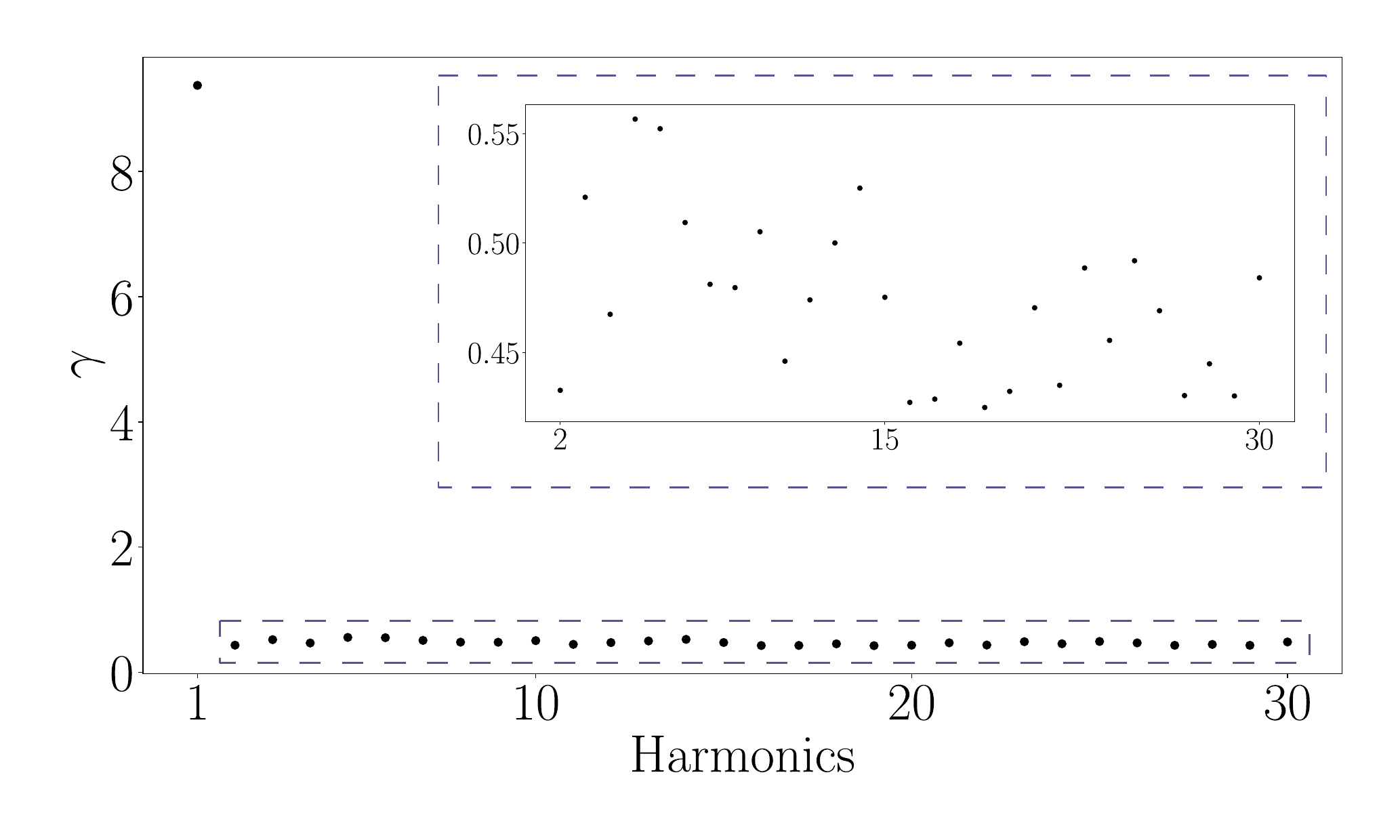}
    \caption{ Coupling term $\gamma$, defined from Eq.~\ref*{eq:gamma_coupl}, obtained from fitting the $\delta \varepsilon$ vs $\varepsilon_0$ maps. As expected, $\gamma$ is an order of magnitude larger for the first harmonic because of the presence of the resonator, while $\gamma \sim 0.5$ for N$>$1 (shown in the inset). This value corresponds to an AC lever arm $\alpha^{rf}_{DTR} \approx 0.35$.}
    \label{fig:gamma_coupling}
\end{figure}

The resulting attenuation deduced from the fits in Fig.~\ref*{fig2}g-h is $0.7 \pm 0.1$ dB.
After this calibration was known, we considered the case of the DTR, where coupling between the gates leads us to define an \textit{effective} rf lever arm

\begin{equation}
    \alpha^{rf}_\text{DTR} = \alpha_{21}^\text{DC} + \gamma \alpha_{11}^\text{DC}
    \label{eq:gamma_coupl}
\end{equation}
\noindent
with $\gamma$ a coupling term between 0 and 1 for $N>1$ and between 0 and the quality factor of the resonator Q for the first harmonic (Fig.~\ref*{fig:gamma_coupling}). All but the first harmonics show $\gamma \sim 0.5$, which corresponds to $\alpha^{rf}_\text{DTR} \approx 0.35$. We attribute the large discrepancy of the first harmonic to direct cross coupling to the resonator from the input G2 to G1, which results in a larger effective lever-arm.

\section{Lindblad formalism:}
\label{app:LB}

To accurately simulate the behaviour of the quantum systems in the non-linear regime, we developed a model of the DTR and ICT based on the Lindblad Master Equation (LME). In this context, the evolution of the density matrix is:

\begin{equation}
\hbar \frac{d}{dt} \rho = -i [H(t), \rho] + \sum_{i = +,-} \hbar \Gamma_i \mathcal{D}[L_i]
\label{eq:LME}
\end{equation}
\noindent
where 
\begin{equation}
    \mathcal{D}[L] = L_i^\dagger \rho L_i - \frac{1}{2} \left\{ L_i^\dagger L_i, \rho \right\}
\end{equation}
\noindent
describes the non-Hamiltonian part of the system evolution and its relaxation towards thermal equilibrium. These dynamics are fully described by a series of \textit{jump operators} $L_i$ and the relative coupling rates $\Gamma_i$. For the DTR, the Hamiltonian is:

\begin{equation}
    H(t) = \frac{1}{2}\begin{pmatrix}
        \varepsilon(t) & 0 \\
        0 & -\varepsilon(t)
    \end{pmatrix}
    \label{eq:DTR_H}
\end{equation}
\noindent
and relaxation occurs via the electron reservoir, through an interaction Hamiltonian of the form:
\begin{equation}
    H_I = \hbar \Gamma_\text{R} (d^\dagger r + r^\dagger d)
    \label{eq:DTR_HI}
\end{equation}
\noindent
where $d$ ($r$) is the fermionic destruction operator for an electron in the QD (reservoir).
Eq.~(\ref{eq:DTR_HI}) gives rise to the jump operators: 

\begin{align}
& L_+ = d^\dagger = \begin{pmatrix}
    0 & 1 \\
    0 & 0
\end{pmatrix}\\
& L_- = d = \begin{pmatrix}
    0 & 0 \\
    1 & 0
\end{pmatrix}
\end{align}
\noindent
with the relevant rates being:

\begin{align}
& \Gamma_+ = \Gamma_\text{R} f(\varepsilon)\\
& \Gamma_- = \Gamma_\text{R} (1- f(\varepsilon))
\end{align}
\noindent
where $f(\varepsilon)$ is the Fermi-Dirac distribution at the QD detuning.
We shall note that, in the particular case of the DTR, the LME can be proven to be formally identical to the well-known semiclassical Master Equation \cite{Cochrane2022_2}
\begin{equation}
    \dot{P} + \Gamma_\text{R} P =  \Gamma_\text{R} f(\varepsilon(t))
    \label{eq:Master_semicl}
\end{equation}
\noindent
whose analytical solution is discussed in Appendix~\ref{app:Y}.

For the ICT, the presence of a tunnel coupling $\Delta$ in the DQD can be described by the Hamiltonian
\begin{equation}
    H(t) = \frac{1}{2}\begin{pmatrix}
        \varepsilon(t) & \Delta \\
        \Delta & -\varepsilon(t)
    \end{pmatrix}
    \label{eq:ICT_H}
\end{equation}
\noindent
In this case, the dominant relaxation process is phonon absorption and emission, which couple to the electronic system via 
\begin{equation}
    H_I = \hbar \Gamma_\text{P} (U d^\dagger U^\dagger a + a^\dagger U^\dagger d U)
    \label{eq:ICT_HI}
\end{equation}

where $a$ is the bosonic destruction operator of the phonon bath. In this model, we assume that phonon relaxation occurs between eigenstates of $H(t)$, hence we introduce the rotation matrix $U$, defined such that $U^\dagger H(t) U$ is always diagonal with eigenvalues
\begin{equation}
    E_{\pm}(t) = \pm \frac{1}{2} \sqrt{\varepsilon(t)^2 + \Delta^2},
    \label{eq:ICT_eigenvals}
\end{equation}
\noindent
meaning that the energy different between ground and excited states is $E = E_+ - E_- = \sqrt{\varepsilon(t)^2 + \Delta^2}$.

Eq.~(\ref{eq:ICT_HI}) gives rise to the jump operators 

\begin{align}
& L_+ = U d^\dagger U^\dagger \\
& L_- = U^\dagger d U
\end{align}
\noindent
with the relevant rates being:

\begin{align}
& \Gamma_+ = \Gamma_\text{P} b(E)\\
& \Gamma_- = \Gamma_\text{P} \left[b(E) + 1\right]
\end{align}
\noindent
where with $b(E) = \langle a^\dagger a \rangle = \frac{1}{e^{E/k_\text{B}T} - 1}$ we indicate the Bose-Einstein distribution at the DQD energy splitting.
It is also possible to include dephasing in this model with a rate $\Gamma_{\phi}$ through the additional jump operator 
\begin{equation}
L_\phi = U \sigma_z U^\dagger.
\end{equation}

The effects of $\Gamma_{\phi}$ are to suppress coherent phenomena, such as Landau-Zener-Stueckleberg-Majorana interferometry, which are not discussed in this work. Numerically iterating Eq.~(\ref{eq:LME}), allows for the simulation of $\rho(t)$, whose Fourier Transform of the diagonal entries as a function of detuning are proportional to the rf reflectometry line-shapes.

\section{DC Lever arm extraction:}
\label{app:alpha}

\begin{figure}[ht]
    \centering
    \includegraphics[width=1\linewidth]{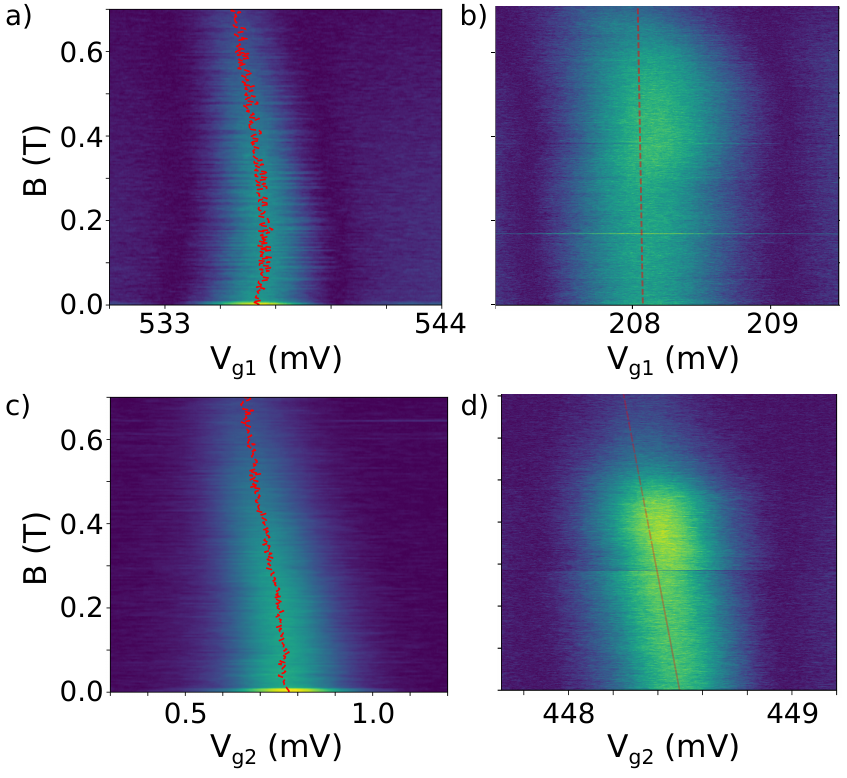}
    \caption{Magnetospectroscopy of dot to reservoir transition used to extract a) $\alpha_{11}$, b) $\alpha_{12}$, c) $\alpha_{21}$ and d) $\alpha_{22}$ respectively. }
    \label{fig:alpha}
\end{figure}

To determine the various lever-arms of our DQD system, we carried out magnetospectroscopy on DTR lines on QDs one and two while sweeping $V_\text{G1}$ and $V_\text{G2}$ accordingly (as shown in Fig~\ref{fig:alpha}). By assuming a $g$ factor of two, we are then able to extract the lever arms $\alpha_{ij}$ as follows
\begin{equation}
    g \mu_B \Delta B = \alpha_{ij}e\Delta V_i,
\end{equation}
\noindent
where $\mu_\text{B}$ is the Bohr magneton, $\Delta B$ is the change in the applied magnetic field and $e$ is the charge of an electron. The extracted lever arm matrix is
\begin{equation}
    \begin{pmatrix}
    \alpha_{11} & \alpha_{12} \\
    \alpha_{21} & \alpha_{22}
    \end{pmatrix} =
    \begin{pmatrix}
        0.59\pm0.02 & 0.022\pm0.006\\
        0.066\pm0.001 & 0.33\pm0.01
    \end{pmatrix}
\end{equation}

For the DTR simulations of QD 1, $\alpha_{11}$ was used, while for the ICT $\alpha_\text{ICT}=\alpha_{22}-\alpha_{21}=0.264\pm0.011$. 

\section{Electron temperature:}
\label{app:Te}
\begin{figure}[ht]
    \centering
    \includegraphics[width=\linewidth]{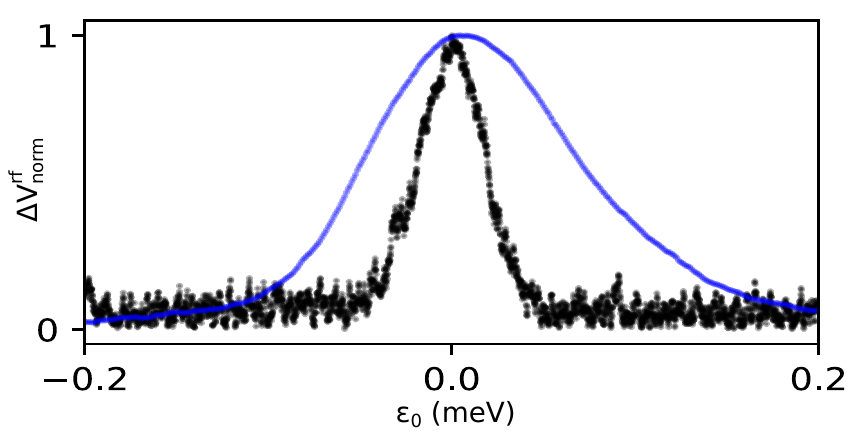}
    \caption{Experimentally measured DTR line-shapes in the thermally (black) and lifetime (blue) broadened regimes, normalized to their maximum value.}
    \label{fig:broadening}
\end{figure}

To set an upper bound for the electron temperature $T_e$, we find the narrowest DTR (black line in Fig.~\ref{fig:broadening}) on QD one and measure its FWHM as a function of the mixing chamber fridge-temperature $T_{\text{fridge}}$. We tune the device to a different region in voltage space, where the signal is narrower than the lifetime broadened DTR used for multiplication (see Fig.~\ref{fig:broadening}).

\begin{figure}[ht]
    \centering
    \includegraphics[width=\linewidth]{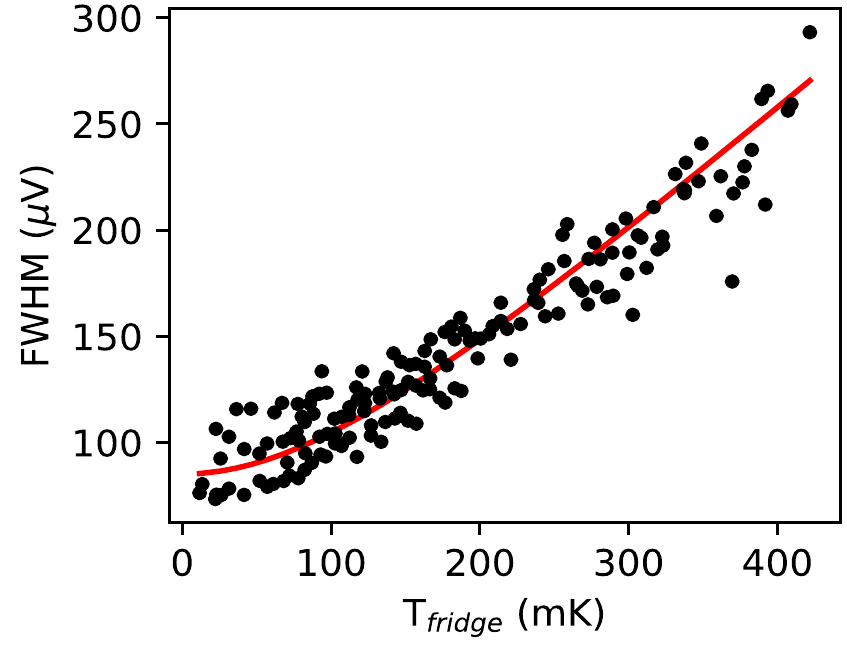}
    \caption{Temperature dependence on the FWHM of a DTR on G1, from which an electron temperature $T_e$ of 140$\pm$3 mK was extracted.}
    \label{fig:Te}
\end{figure}

We ensured no power broadening by measuring the transition at varying rf powers~\cite{Ahmed2018}. From the temperature dependence measurements in Fig.~\ref{fig:Te}, we fit the following expression to extract $T_e$,
\begin{equation}
    \text{FWHM} = \frac{3.53 k_B}{e \alpha_{11}}\sqrt{T^2_{\text{fridge}}+T^2_e}.
\end{equation}

From the fit, we estimate an upper bound for the electron temperature of 140$\pm$3 mK.

\newpage
\section{DTR harmonics:}

\begin{figure*}[ht]
    \centering
    \includegraphics[width=0.9\linewidth]{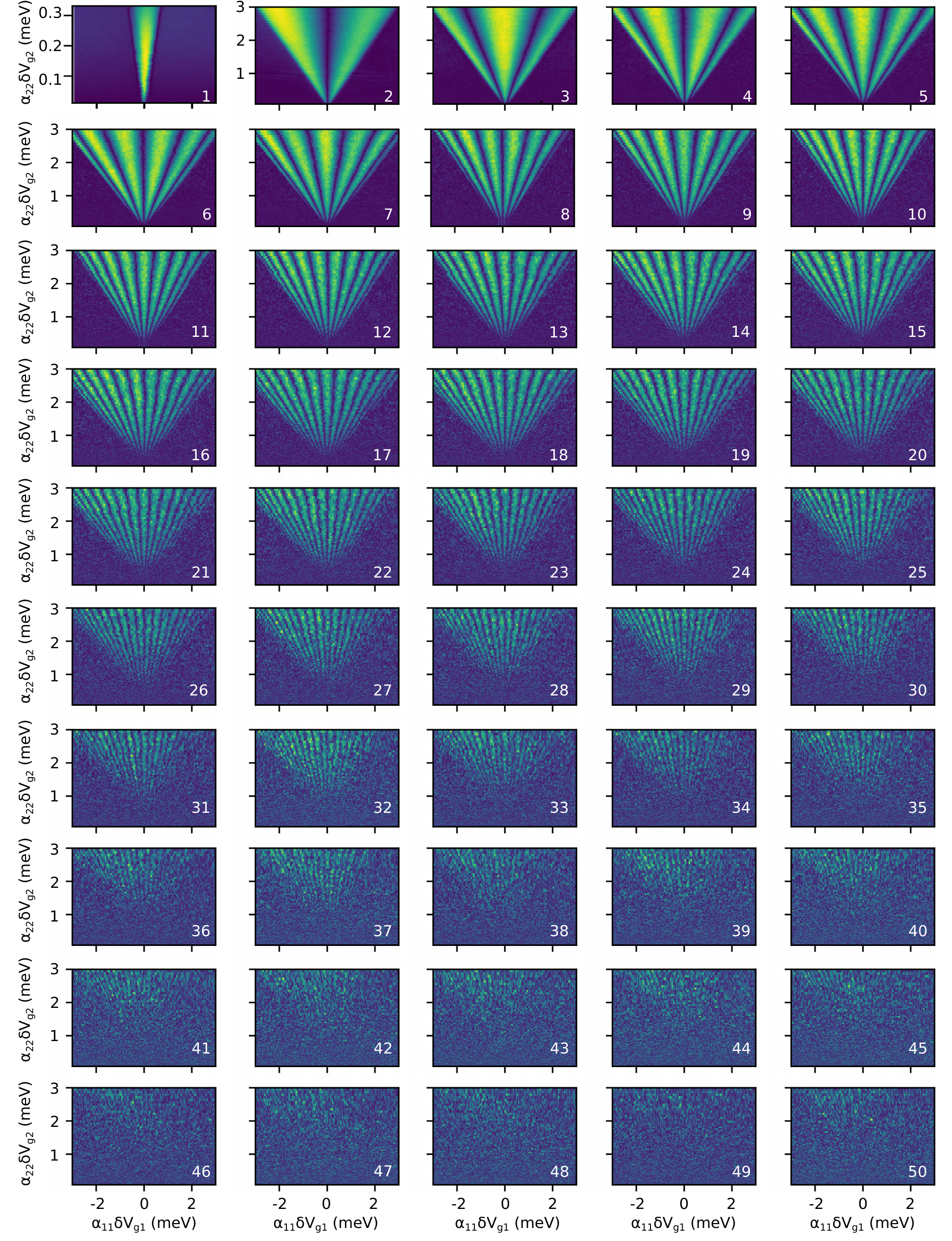}
    \caption{Raw data of first 50 harmonics experimentally measured across a DTR}
    \label{fig:DTR}
\end{figure*}

\begin{figure*}[ht]
    \centering
    \includegraphics[width=0.9\linewidth]{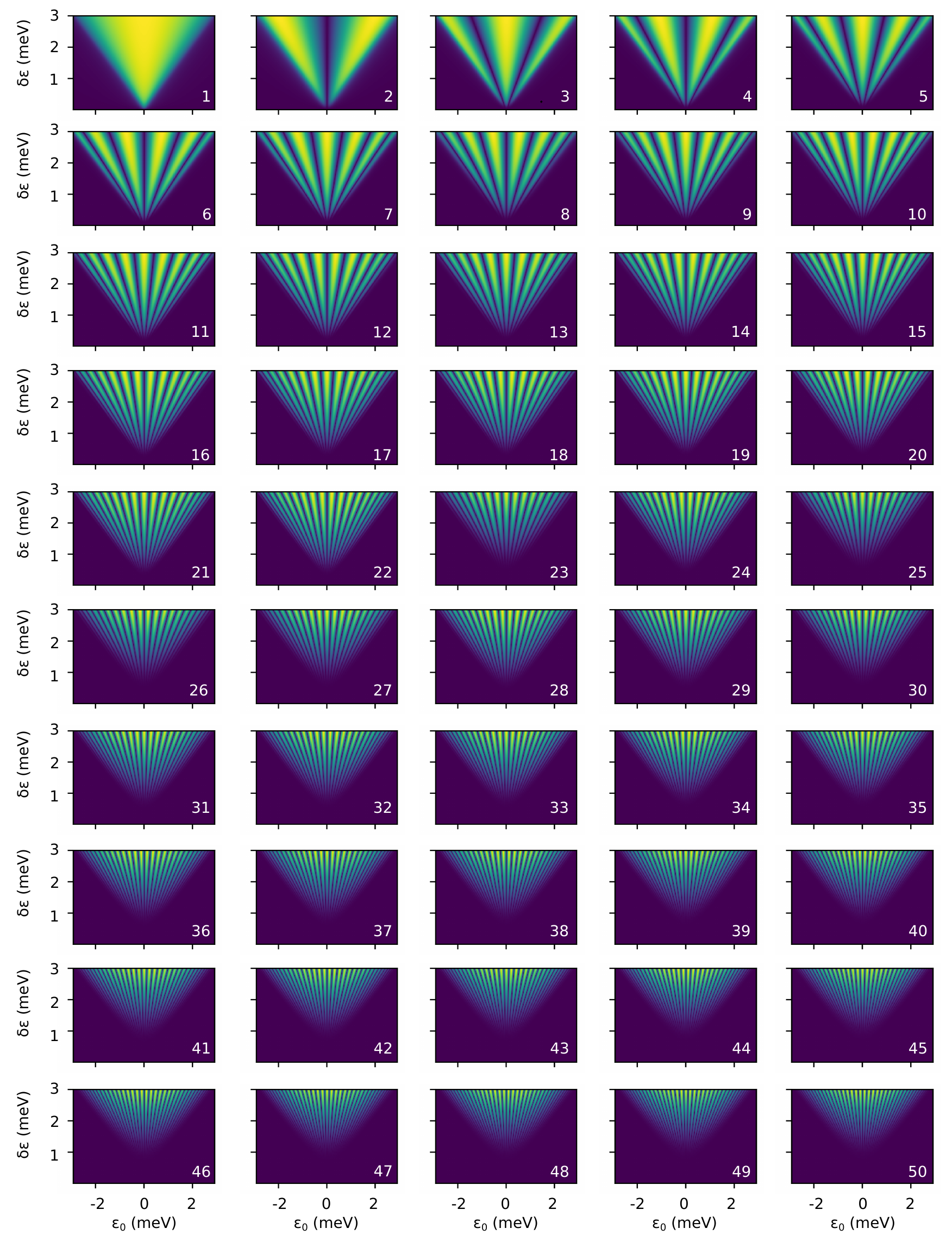}
    \caption{Simulated data of first 50 DTR harmonics}
    \label{fig:DTR_fit}
\end{figure*}

In this section, we present measurements and simulations of the DTR harmonics from $N=1$ to $50$, see Fig.~\ref{fig:DTR} and ~\ref{fig:DTR_fit}, respectively. 
\label{app:DTR_50}

\section{ICT harmonics:}

\begin{figure*}[ht]
    \centering
    \includegraphics[width=0.9\linewidth]{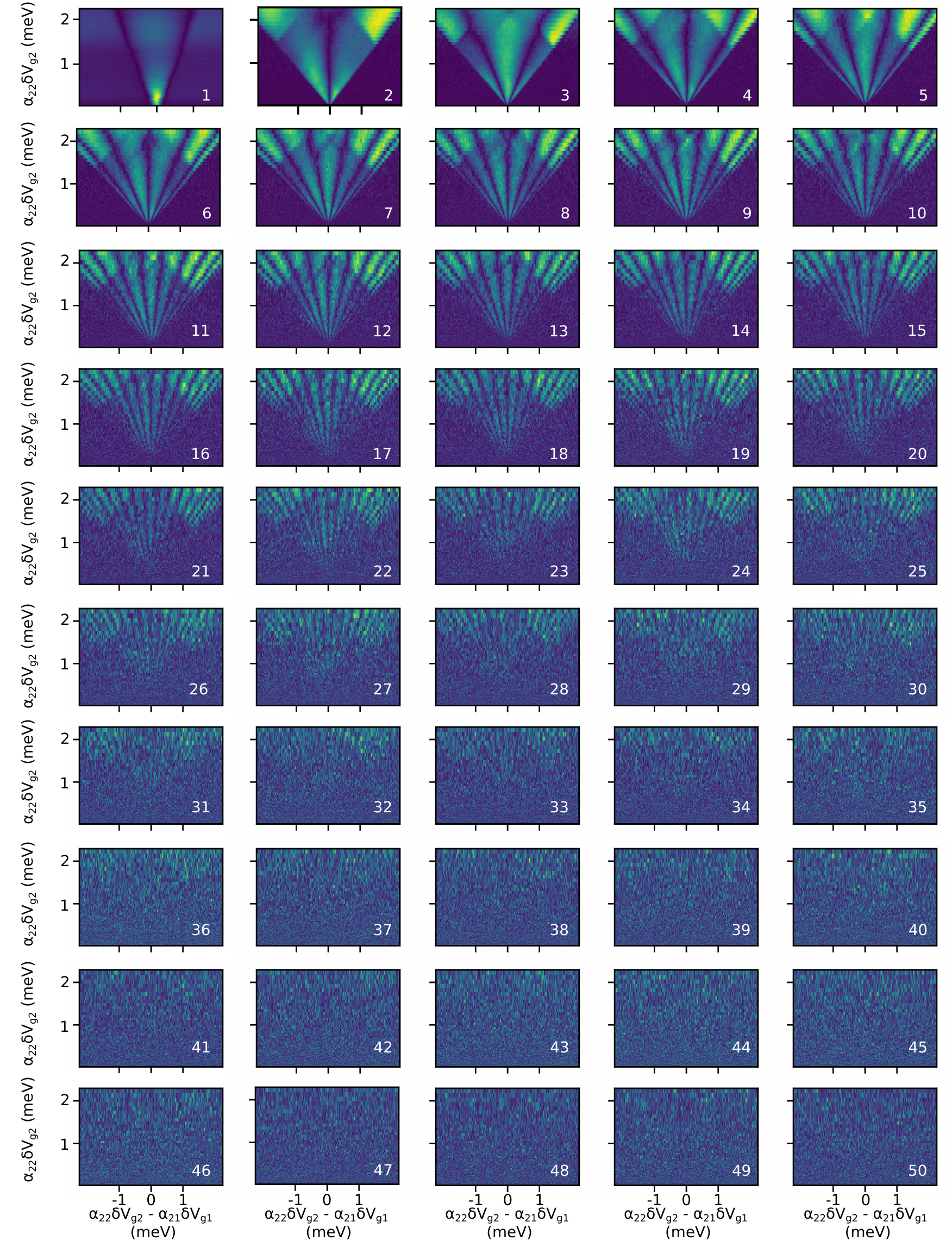}
    \caption{Raw data of first 50 harmonics experimentally measured across an ICT}
    \label{fig:ICT}
\end{figure*}

\begin{figure*}[ht]
    \centering
    \includegraphics[width=0.9\linewidth]{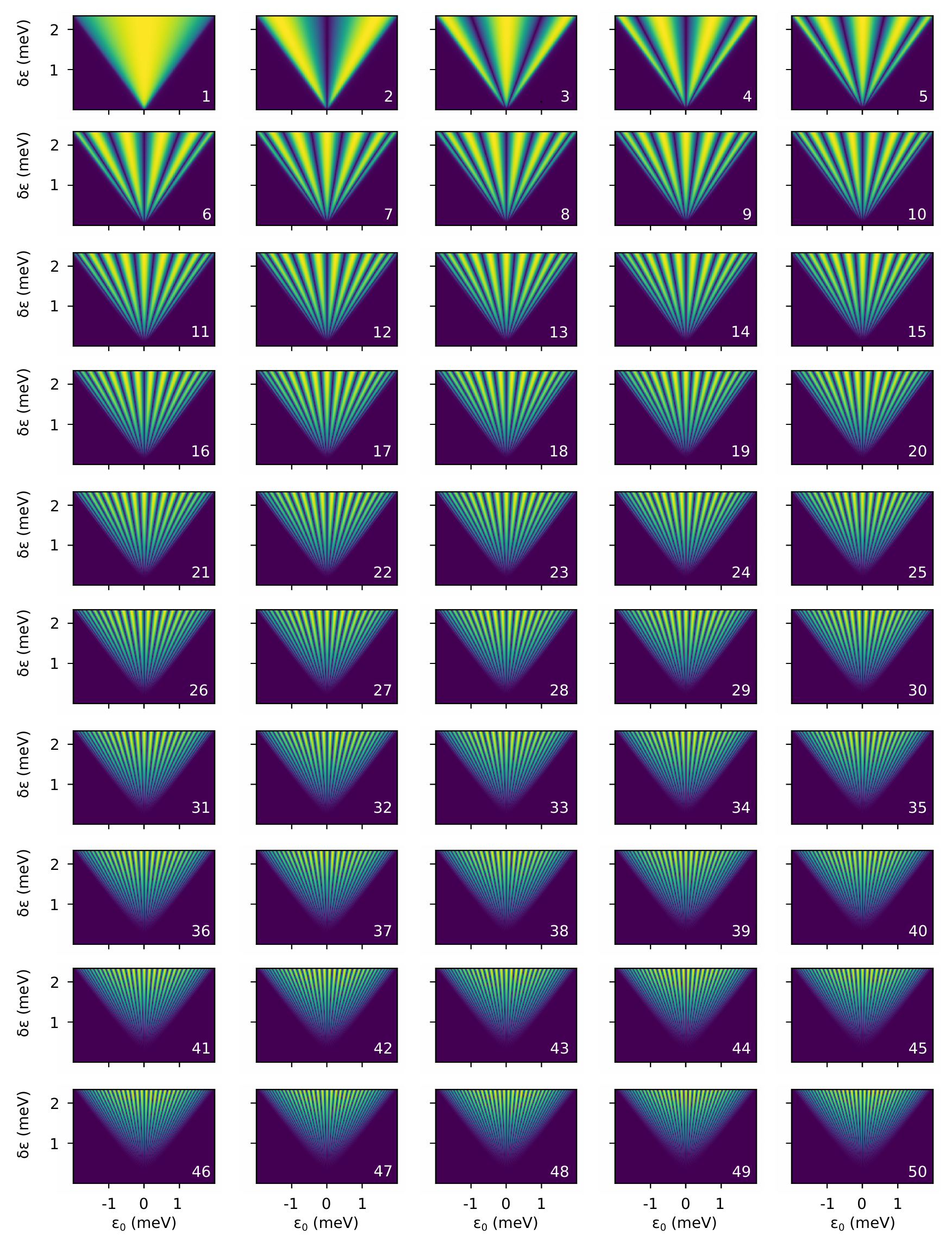}
    \caption{Simulated data of first 50 ICT harmonics}
    \label{fig:ICT_fit}
\end{figure*}

In this section, we present measurements and simulations of the ICT harmonics from $N=1$ to $50$, see Fig.~\ref{fig:ICT} and~\ref{fig:ICT_fit}, respectively.

\label{app:ICT_50}

\section{Power conversion efficiency extraction:}
\label{app:power_conversion}

\begin{figure}[ht]
    \centering
    \includegraphics[width=1\linewidth]{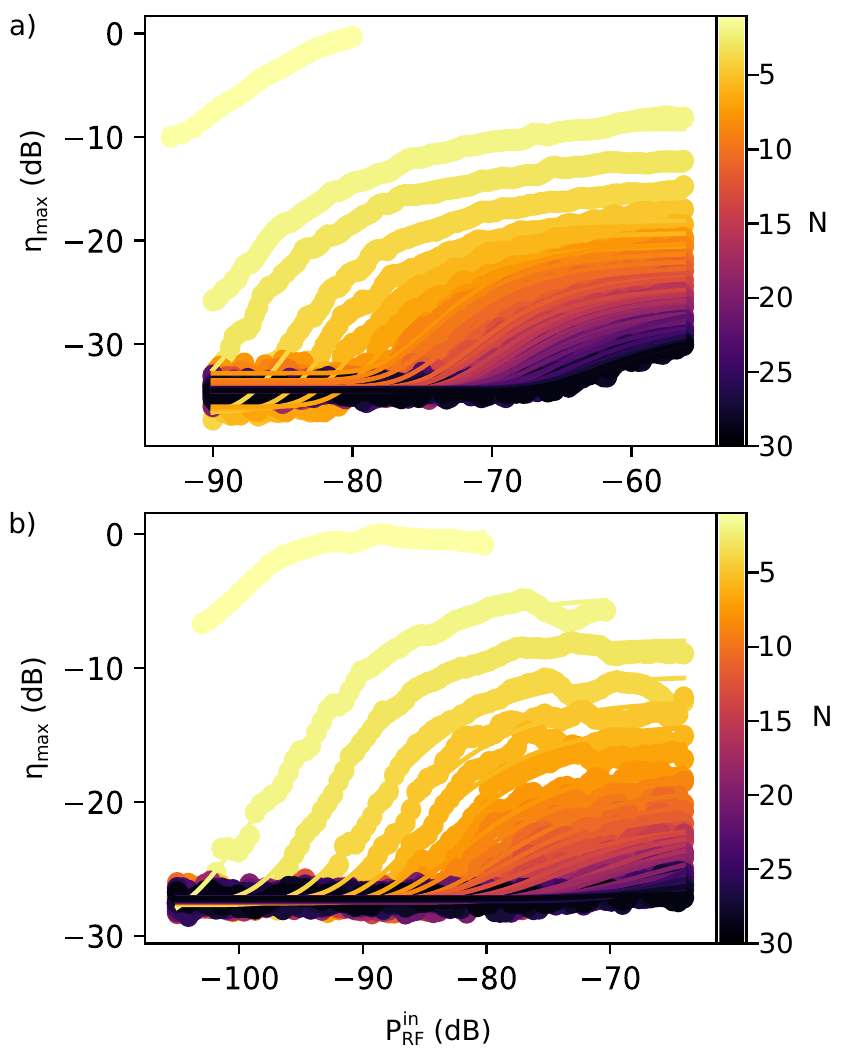}
    \caption{Power conversion measured for the first 30 harmonics for the DTR (a) and ICT (b), which were used to extract the data in Fig.~\ref{fig4}a.}
    \label{fig:power_conversion}
\end{figure}

The measurements in Appendices~\ref{app:DTR_50} and~\ref{app:ICT_50} were used to extract the power conversion efficiency for the DTR and ICT.  
The acquired data was limited to rf powers where no excited states or interference between multiple transitions were present (see Appendix~\ref{app:n**2}). The maximum signal would then occur at zero detuning for the odd harmonics, while for the even ones, it depends linearly on $\delta \epsilon$. For each harmonic, we have extracted the measured maximum signal at each rf power along with the corresponding fitted value, as shown in Fig.~\ref{fig:power_conversion}. These were then used to plot the power conversion relationship shown in Fig.~\ref{fig4}a.   

\section{Beyond N\texorpdfstring{\pdfmath{^{-2}}}{-2} dependence:}
\label{app:n**2}

\begin{figure*}[ht]
    \centering
    \includegraphics[width=1\linewidth]{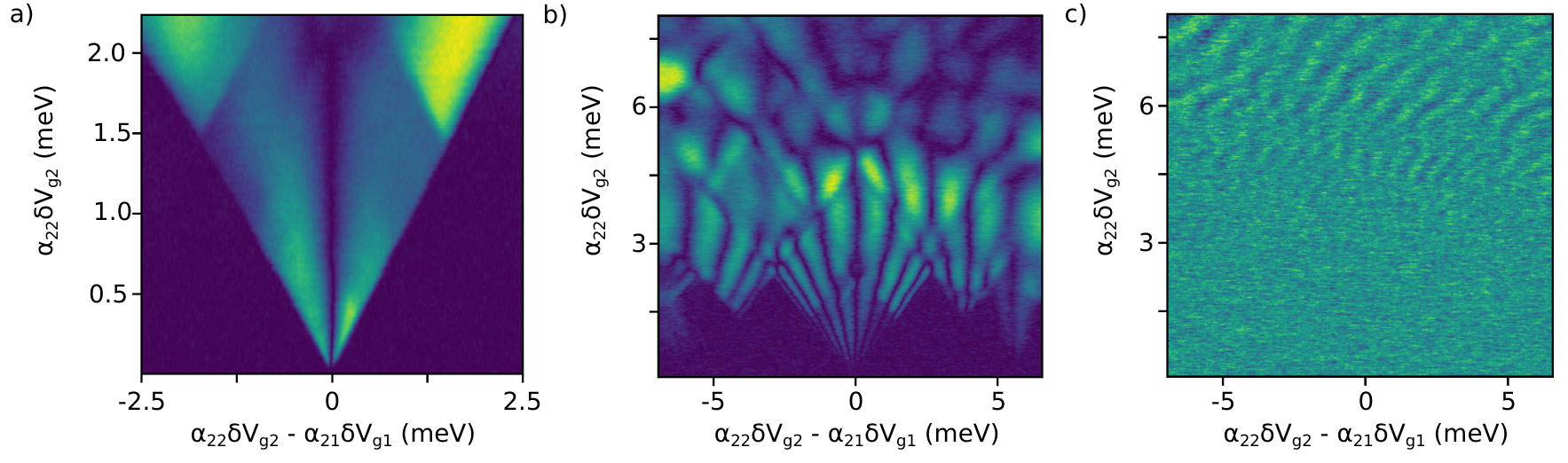}
    \caption{Increasing power conversion above the expected N$^{-2}$ dependence. a) Second harmonic ICT transition highlighting how an increase in power conversion is obtainable due to the presence of excited states above a driving detuning $\delta \epsilon\sim$1.5 meV. b) the Tenth harmonic showcasing how multiple ICTs can interfere at high enough detunings. By modulating multiple electrons, an increase in power conversion is possible resulting in regions of constructive interference. c) Measured signal at $\omega_r/75$ obtained from the interference of multiple ICT transitions.}
    \label{fig:n**2}
\end{figure*}

At high powers, it is possible to drive the system strongly to overcome the expected power conversion dependence of $N^{-2}$. In particular, for the ICT, an increment in signal is possible if excited states are present, as shown in Fig.~\ref{fig:n**2}a. This is due to a gain in quantum capacitance due to an increase in curvature of the eigenenergies. By driving the system even harder, it is possible to get multiple charges $n$ to transition, resulting in an interference pattern, as highlighted in Fig.~\ref{fig:n**2}b. We can thus find regions in voltage that should theoretically result in an increase of $n$ in power conversion that will counteract the expected $N^{-2}$ relationship; this allowed us to measure a signal at even higher harmonics, such as $f_\text{r}/75$, as shown in Fig.~\ref{fig:n**2}c.

\section{Ideal phase noise:}
\label{app:phase_noise}

The phase noise of an rf source dominated by $1/f$ noise can be empirically described by Leeson's equation~\cite{lee2000oscillator}
\begin{equation}
    \label{eq:Leeson}
    \mathcal{L}_p(\omega_0, \Delta \omega) = 10 \text{log} \Bigg[ \frac{2 F k_B T}{P_{\text{sig}}} \Big(1 + \frac{\omega_0}{2 Q \Delta \omega}\Big)^2 \Big(1 + \frac{\Delta \omega_{1/f^3}}{|\Delta \omega|}\Big)\Bigg],
\end{equation}
\noindent
where $\Delta \omega$ is the offset frequency from the carrier frequency $\omega_0$, $F$ is an empirically fitted parameter known as the noise factor, $P_{\text{sig}}$ is the power of the rf source, $Q$ is the loaded quality factor, and $\Delta \omega_{1/f^3}$ is the $1/f$ corner frequency. The $\frac{\omega_r}{2 Q \Delta \omega}$ dependence arises from the $1/f$ decrease in voltage frequency response for an ideal $RLC$ tank circuit as a function of $\Delta \omega$, which is then squared to convert it into power. The noise factor $F$ and the additive one within the first set of brackets account for the noise floor. The second parentheses provide the 1/$|\Delta \omega|^3$ behaviour observed at limited offset frequencies. 

An ideal frequency multiplier performs a rigid shift in the frequency axis, which, after multiplication, increases the phase noise by a factor of $N$. For harmonics at the same frequency offset $\Delta \omega$, as is the case in our data in Fig.~\ref{fig4}, the ideal phase noise is:

\begin{equation}
    \Delta \mathcal{L}_p^{N, \text{ideal}} = \mathcal{L}_p \left(\frac{\omega_r}{N}, \frac{\Delta \omega}{N}\right)  - \mathcal{L}_p \left(\frac{\omega_r}{N}, \Delta \omega\right)
    \label{eq:ideal_phase_noise}
\end{equation}
\noindent

\bibliography{bibliography}


\end{document}